\documentclass[aps,prb,twocolumn,showpacs,groupedaddress,amsmath]{revtex4}

\usepackage{graphicx}
\usepackage{bm}
\usepackage{here}

\newcommand{\bra}[1]{\langle #1|}
\newcommand{\ket}[1]{|#1\rangle}
\newcommand{\braket}[2]{\langle #1|#2\rangle}

\begin{document}

\title{Quasiparticle spectral weights of Gutzwiller-projected high $T_c$ superconductors}

\author{Samuel Bieri}
\email[]{samuel.bieri@epfl.ch}

\author{Dmitri Ivanov}

\affiliation{ Institute of Theoretical Physics, Ecole Polytechnique
F\'{e}d\'{e}rale de Lausanne (EPFL), CH-1015 Lausanne, Switzerland}

\pacs{74.72.-h, 71.10.Li, 71.18.+y}


\date{August 2, 2006}

\begin{abstract}
We analyze the electronic Green's functions in the superconducting
ground state of the $t$--$J$ model using Gutzwiller-projected wave
functions, and compare them to the conventional BCS form. Some of
the properties of the BCS state are preserved by the projection: the
total spectral weight is continuous around the quasiparticle node
and approximately constant along the Fermi surface. On the other
hand, the overall spectral weight is reduced by the projection with
a momentum-dependent renormalization, and the projection produces
electron-hole asymmetry in renormalization of the electron and hole
spectral weights. The latter asymmetry leads to the bending of the
effective Fermi surface which we define as the locus of equal
electron and hole spectral weight.
\end{abstract}

\maketitle

\section{Introduction}

Shortly after the discovery of superconductivity in copper-oxyde
compounds,\cite{bednorz86} Anderson proposed a Gutzwiller-projected
BCS wave function which would describe the superconducting ground
state of high-temperature superconductors.\cite{anderson87} The
variational approach to superconducting cuprates based on Anderson's
original proposal has since had a lot of success, while the strong
Coulomb repulsion and the non-perturbative nature of the problem
make other approaches extremely difficult. Interest in projected
wave functions as variational ground states for cuprate
superconductors was initiated by several research groups in late
80's \cite{gros88, yokoyama88} and lead to a considerable activity
in the field. The projected wave functions show large overlap with
exact ground states on small clusters and have low variational
energies for the $t\text{--}J$ model.\cite{hasegawa89,becca00}
Furthermore, several experimental properties of cuprates like the
zero-temperature phase diagram and d-wave pairing symmetry are
extremely well predicted within this approach.\cite{paramekanti0103,
rmft}

Due to considerable progress of the experimental technique of Angle
Resolved Photoemission Spectroscopy (ARPES) on cuprates, more and
more high-quality data on the low-lying spectral properties of these
compounds have been made available in recent years.\cite{ARPES}
Experimentally, the low-energy excitations of superconducting
cuprates are known to resemble BCS quasiparticles
(QPs).\cite{matsui03} It is therefore interesting to theoretically
explore the wave function of projected QP excitations and compare
them to unprojected BCS QPs. The most apparent differences are the
the doping dependency of the nodal Fermi velocity and the
renormalizations of the nodal QP spectral weight and of the current
carried by QPs.\cite{paramekanti0103, yunoki05, yunoki05prl,
edegger06, nave06} In the present paper we further analyze the
properties of the superconducting ground state and the QP
excitations with the Variational Monte Carlo technique
(VMC).\cite{gros78} We calculate the equal-time Green's functions,
both normal and anomalous, in the Gutzwiller-projected state and
derive from them the QP spectral weights for addition and removal of
an electron at zero temperature. The main conclusion of our study is
that, due to a non-trivial interplay of superconductivity and strong
Coulomb repulsion (projection), the electron and hole spectral
weights are renormalized differently. A natural way to describe this
asymmetry is to define the ``effective Fermi surface'' as the locus
of points where the electron and the hole spectral weights are
equal. Thus defined Fermi surface acquires an additional outward
bending in the anti-nodal region as compared to the original
unprojected Fermi surface. This bending is a signature of a
deviation from the BCS theory and may be responsible for the shape
of the Fermi surface observed in ARPES experiments. The validity of
Luttinger's rule\cite{luttinger61} in strongly interacting and
superconducting materials has been questioned experimentally and
theoretically recently.\cite{ARPES,sensarma06,gros06} Our findings
provide further indication of its inapplicability in strongly
correlated superconductors.

The paper is organized as follows. Section \ref{sec:model} contains
the definition of the model and wave functions used in our
calculations. In Section \ref{sec:weights} we first present some
exact relations for projected wave functions, we then describe our
results on the QP spectral weights. Section \ref{sec:sc} is devoted
to the calculation of the equal-time anomalous Green's function.
Finally, Section \ref{sec:FS} defines the ``effective Fermi
surface'' and discusses its deviation from the unprojected one.

\section{The model}
\label{sec:model}

In the tight-binding description, the cuprates are modeled by
electrons hopping on a square lattice. The appropriate model is the
$t$--$J$ Hamiltonian:
\begin{equation} \label{eq:tJ}
H_{t\text{--}J} = -t\sum_{<i,j>,\sigma} P_G\, c_{i\sigma}^\dagger
c_{j\sigma}\, P_G + J \sum_{<i,j>} ({\bm S}_i\cdot{\bm S}_j -
\frac{n_i n_j}{4})
\end{equation}
acting in the Hilbert-space with less than two electrons per site.
Here $n_{i\sigma} = c_{i\sigma}^\dagger c_{i\sigma}$, ${\bm S}_i =
\frac{1}{2}c_{i\sigma}^\dagger \bm{\sigma}_{\sigma\sigma'}
c_{i\sigma'}$, $c^\dagger_{i\sigma}$ is the electron creation
operator in the Wannier state at site $i$, and $\bm \sigma$ are the
Pauli matrices. The no-double occupancy is preserved by the
Gutzwiller projector $P_G = \Pi_i(1-n_{i\uparrow}n_{i\downarrow})$.

The $t$--$J$ model can be viewed as the large $U$ limit of the
one-band Hubbard model, neglecting the 3-site-hopping term. Provided
that the model is analytic in $t/U$, doubly occupied sites can be
re-introduced perturbatively to recover the full Hilbert space of
the Hubbard model.\cite{fazekas99, paramekanti0103} Although the
inclusion of these corrections present no major difficulty, we chose
to neglect them here. In most quantities, only small corrections
arise from finite double occupancy,\cite{paramekanti0103,nave06}
which makes this approach to the large-$U$ Hubbard model consistent.
Furthermore, it has been argued that the $t$--$J$ model is in fact
more appropriate than the one-band Hubbard model in describing the
CuO planes.\cite{zhang88}

We consider the usual variational ground state:\cite{gros88}
\begin{equation}\label{eq:wf}
\ket{\Psi_{H}} =  P_H P_G \ket{dBCS(\Delta,\mu)}
\end{equation}
where $P_H$ is the particle number projector on the subspace of
$L^2-H$ electrons; $L^2$ is the total number of sites. Both hole
number and number of sites are even. $\ket{dBCS}$ is the ground
state of the BCS mean field Hamiltonian with nearest neighbor
hopping and d-wave pairing symmetry on the square lattice:
$\ket{dBCS} = \Pi_{\bm{k}} (u_{\bm{k}} + v_{\bm{k}}
c_{\bm{k}\uparrow}^\dagger c_{-\bm{k}\downarrow}^\dagger)\ket{0}
\propto \Pi_{\bm{k},\sigma}\gamma_{\bm{k}\sigma}\ket{0}$.
$\gamma_{\bm{k}\sigma} = u_{\bm{k}} c_{\bm{k}\sigma} + \sigma
v_{\bm{k}} c_{-\bm{k}\bar\sigma}^\dagger$, $u_{\bm{k}}^2 =
\frac{1}{2}(1-\frac{ \xi_{\bm{k}} }{E_{\bm{k}}})$, $v_{\bm{k}}^2 =
\frac{1}{2}(1+\frac{\xi_{\bm{k}} }{E_{\bm{k}}})$, $E_{\bm{k}} =
\sqrt{\xi_{\bm{k}}^2 + \Delta_{\bm{k}}^2}$, $\xi_{\bm{k}} = - 2
(\cos(k_x) + \cos(k_y)) - \mu$, $\Delta_{\bm{k}} = \Delta(\cos(k_x)
- \cos(k_y))$. The wave function \eqref{eq:wf} has two free
parameters: $\Delta$ and $\mu$. These variational parameters are
chosen to minimize the energy of the $t$--$J$ Hamiltonian
\eqref{eq:tJ} for the experimentally relevant value $J/t = 0.3$ and
for every doping level. We use the optimized parameters and the
cluster geometry from Ref.~\onlinecite{ivanov03}.

The following ansatz is used for the excited states:\cite{rmft,
paramekanti0103, yunoki05, yunoki05prl, edegger06, nave06}
\begin{equation}\label{eq:excit}
\ket{\Psi_{H,\bm{k},\sigma}} = P_H P_G
\gamma^\dagger_{\bm{k}\sigma}\ket{dBCS}
\end{equation}
In the following, the normalized versions of \eqref{eq:wf} and
\eqref{eq:excit} will be denoted by $\ket{H}$ and
$\ket{H,\bm{k},\sigma}$, respectively:
\begin{subequations}\begin{eqnarray}
\ket{H} &=& \Vert \Psi_{H} \Vert^{-1} \ket{\Psi_{H}} \\
\ket{H,\bm{k},\sigma} &=& \Vert \Psi_{H,\bm{k},\sigma} \Vert^{-1}
\ket{\Psi_{H,\bm{k},\sigma}}
\end{eqnarray}\end{subequations}

The many-particle wave function \eqref{eq:wf}, sometimes called
Anderson's Resonant Valence Bond (RVB) wave function, implements
both strong electron correlations and superconductivity. It is known
to have a considerable overlap with the true ground state of the
$t$--$J$ model at non-zero hole doping on small
clusters.\cite{becca00, park05, poilblanc94, hasegawa89} There is
also numerical support from exact diagonalization studies indicating
well defined BCS-like QPs as low-energy excitations of the $t$--$J$
model.\cite{ohta94} Therefore, the excited trial states
\eqref{eq:excit} are expected to be close to the true excitations of
the $t$--$J$ model. However, here we are more interested in the
physical content of the proposed wave functions than in their
closeness to the eigenstates of a particular
Hamiltonian.\cite{note_norm}

\section{Quasiparticle spectral weights}
\label{sec:weights}

The QP spectral weights are defined as the overlap between the bare
electron/hole added ground state and the QP excitations of the
model:
\begin{equation}
Z^{\pm}_{\bm k, H} = |\bra{H\mp 1, \bm k,\sigma} c_{\bm
k,\sigma}^{\pm} \ket{H}|^2
\end{equation}
where $c^{\pm}_{\bm k,\sigma}$ are the bare electron
creation/anihilation operators.

It is well-known that the QP spectral weight for adding an electron
(equal-time normal Green's function) can be calculated from the
ground state spectral function:\cite{yunoki05}
\begin{equation} \label{eq:zp}
Z^{+}_{\bm k, H} = \frac{1+x}{2} -  \langle c^{\dagger}_{\bm
k\sigma} c_{\bm k\sigma}\rangle
\end{equation}
where $x = H/L^2$ is the hole concentration.

The QP spectral weight for adding a hole is more difficult to
calculate. It is useful to note that it can also be calculated from
ground state expectation values,
\begin{equation}\label{eq:zpm}
Z^{+}_{\bm k, H+1} Z^{-}_{\bm k, H-1} = \lvert \Phi_{\bm k,H}
\rvert^2\, ,
\end{equation}
where $\Phi_{\bm k,H}$ is the superconducting order parameter
(equal-time anomalous Green's function)
\begin{equation}\label{eq:cc}
\Phi_{\bm k,H} = \bra{H+1} c_{\bm k\uparrow} c_{-\bm k\downarrow}
\ket{H-1}\, .
\end{equation}
Relation (\ref{eq:zpm}) can be proven by algebraic manipulations
with the Gutzwiller projector. It is exact in finite systems and is
also valid in the thermodynamic limit:
\begin{equation}\label{eq:zpm-therm}
Z^{+}_{\bm k} Z^{-}_{\bm k} = \lvert \Phi_{\bm k} \rvert^2\, .
\end{equation}
Remarkably, this relation holds for both the unprojected BCS state
(with $Z^{+}_{\bm k}=|v_{\bm k}|^2$,  $Z^{-}_{\bm k}=|u_{\bm k}|^2$)
and the fully projected wave function. It would be interesting to
explore if this relation is valid in a more general case of
superconducting systems or if it is just a peculiarity of a certain
class of wave functions.

Further, we define the total spectral weight
\begin{equation}
Z^{tot}_{\bm k}= Z^{+}_{\bm k} + Z^{-}_{\bm k}\, .
\end{equation}
The main contribution to $Z^{tot}_{\bm k}$ is given by $Z^{+}_{\bm
k}$ outside the Fermi surface and by $Z^{-}_{\bm k}$ inside. We can
prove an exact upper bound on $Z^{tot}_{\bm k}$:
\begin{equation}
Z^{tot}_{\bm k} < \frac{1+x}{2}\, .
\end{equation}
A proof may be performed by defining the two states
\begin{subequations}\begin{eqnarray}
\ket{a} &=& c_{\bm k\uparrow} \ket{H-1} \\
\ket{b} &=& P_G\, c^\dagger_{-\bm k\downarrow} \ket{H+1}
\end{eqnarray}\end{subequations}
(with this definition, $\ket{b}$ is proportional to
$\ket{H,\bm{k},\uparrow}$). Using \eqref{eq:zp} and
\eqref{eq:zpm-therm}, we show that
\begin{eqnarray}
&&\det \begin{pmatrix}
\braket{a}{a} & \braket{a}{b} \\
\braket{b}{a} & \braket{b}{b}
\end{pmatrix}
\\
&=&\left(\frac{1+x}{2} - Z^{+}_{\bm k} \right) Z^{+}_{\bm k}
-|\Phi_{\bm k}|^2 = Z^{+}_{\bm k} \left( \frac{1+x}{2} -
Z^{tot}_{\bm k} \right) \, . \nonumber
\end{eqnarray}
On the other hand, the same determinant equals
\begin{equation}
\frac{1}{2}\Vert \ket{a}\otimes\ket{b} - \ket{b}\otimes\ket{a} \Vert
^2 >0\, ,
\end{equation}
which completes the proof.

Numerically, we compute the spectral weight $Z^{-}_{\bm k,H-1}$ by
first computing $Z^{+}_{\bm k,H+1}$ and $\Phi_{\bm k,H}$, and then
using (\ref{eq:zpm}). The disadvantage of this method is large error
bars around the center of the Brillouin zone where both $Z^{+}_{\bm
k}$ and $\Phi_{\bm k}$ are small (recently performed calculations of
$Z^{-}_{\bm k}$ by direct sampling of the excited states are free
from this problem).\cite{yunoki06} However, our precision is
sufficient to establish that the total spectral weight $Z^{tot}_{\bm
k}$ is a smooth function and has no singularity at the nodal point.

Technically, $\Phi_{\bm k,H}$ is computed as $|\Phi_{\bm k,H}|^2 =
\Phi_{\bm k,H}^{+} \Phi^{-}_{\bm k,H}$, where
\begin{subequations}\begin{eqnarray} \Phi_{\bm k,H}^{+} &=& \frac{ \langle
\Psi_{H+1}| c_{\bm k\uparrow} c_{-\bm k\downarrow}
|\Psi_{H-1}\rangle }{
\Vert \Psi_{H+1} \Vert^2} \\
\Phi_{\bm k,H}^{-} &=& \frac{ \langle \Psi_{H+1}| c_{\bm k\uparrow}
c_{-\bm k\downarrow} |\Psi_{H-1}\rangle }{ \Vert \Psi_{H-1} \Vert^2}
\, .
\end{eqnarray}\end{subequations}
Both matrix elements can be computed within the usual Metropolis
algorithm.\cite{gros78}

\hrulefill \vspace{.2cm}

\begin{figure}[!h]
\includegraphics[scale=.52]{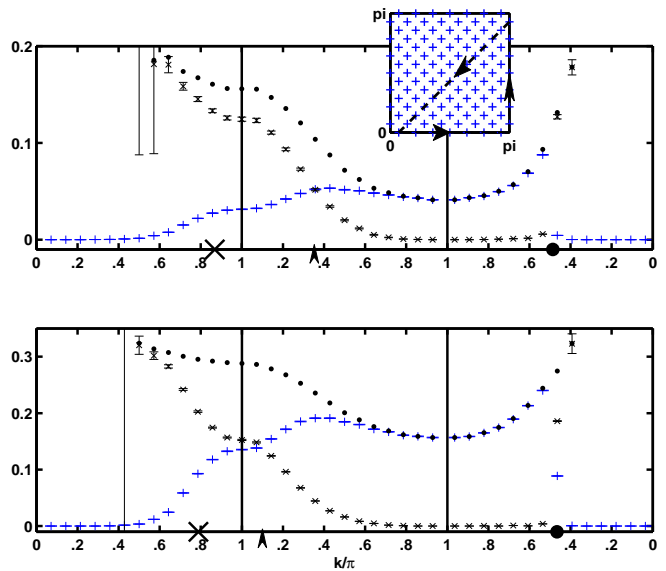}
\caption{(Color online) QP spectral weights for 6 holes (upper plot,
$x\simeq 3\%$) and 22 holes (lower plot, $x\simeq 11\%$) on 196
sites. The spectral weights are plotted along the contour
$0\to(0,\pi)\to(\pi,\pi)\to 0$ (shown in inset). Plus signs ($+$,
blue online) denote the spectral weight $Z_{\bm k}^{+}$, crosses
($\times$) denote $Z_{\bm k}^{-}$, error bars are shown. Solid dots
(black) denote their sum, the total spectral weight $Z_{\bm
k}^{tot}$, error bars not shown. On the horizontal axis, the star
($*$) denotes the intersection with the unprojected Fermi surface
along the $0\to(0,\pi)$ direction; the thick dot is the nodal point.
$Z_{\bm k}^{+}$ and $Z_{\bm k}^{-}$ jump at the nodal point, while
$Z_{\bm k}^{tot}$ is continuous. The intersection with the effective
Fermi surface (see section \ref{sec:FS}) is marked by an arrowhead.
On the diagonal (last segment), $k$ is given in units of $\sqrt{2}$.
\label{fig:cut_z_6_22_196}}
\end{figure}

\begin{figure}[!h]
\includegraphics[scale=.52]{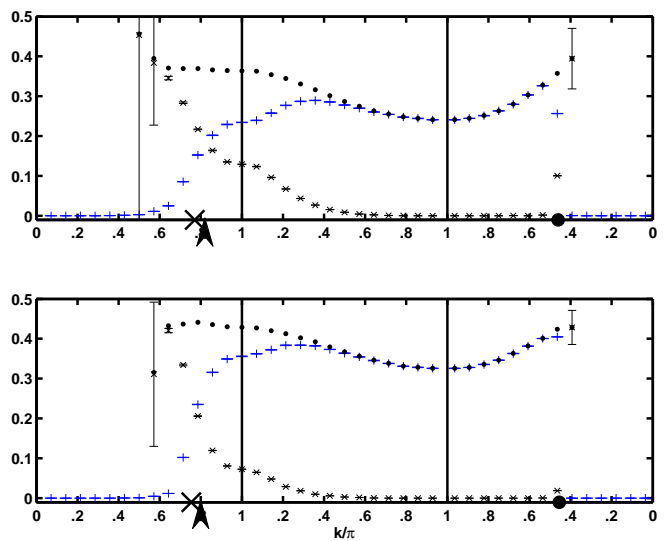}
\caption{(Color online) Same plot as Fig.~\ref{fig:cut_z_6_22_196}
of the QP spectral weights for 34 holes (upper plot, $x\simeq 17\%$)
and 46 holes (lower plot, $x\simeq 23\%$) on 196
sites.\label{fig:cut_z_34_46_196}}
\end{figure}

In Figs.~\ref{fig:cut_z_6_22_196} and \ref{fig:cut_z_34_46_196}, we
plot the spectral weights $Z^+_{\bm k}$, $Z^{-}_{\bm k}$, and
$Z^{tot}_{\bm k}$ along the contour $0\to(0,\pi)\to(\pi,\pi)\to 0$
in the Brillouin zone for different doping levels.
Figure~\ref{fig:nFS} shows the contour plots of $Z^{tot}_{\bm k}$ in
the region of the Brillouin where our method produces small error
bars.\cite{note_comp} From these data, we can make the following
observations:
\begin{itemize}
\item In the case of an unprojected BCS wave function, the total
spectral weight is constant and unity over the Brillouin zone.
Introducing the projection operator, we see that for low doping
($x\simeq3\%$), the spectral weight is reduced by a factor up to
$20$. The renormalization is asymmetric in the sense that the
electronic spectral weight $Z^{+}_{\bm k}$ is more reduced than the
hole spectral weight $Z^{-}_{\bm k}$. For higher doping ($x\simeq
23\%$), the spectral weight reduction is much smaller and the
electron-hole asymmetry decreases.
\item Since there is no electron-hole mixing along the zone
diagonal, the spectral weights $Z^{+}_{\bm k}$ and $Z^{-}_{\bm k}$
have a discontinuity at the nodal point. Our data shows that the
total spectral weight is continuous across the nodal point. Strong
correlations does not affect this feature of uncorrelated BCS
theory. Recently, it has been argued in Ref.~\onlinecite{yang06}
that the total spectral weight of the projected
(non-superconducting) Fermi sea should be continuous across the
Fermi surface. This is consistent with our result.
\end{itemize}

\vspace{.2cm}

The intensities measured in ARPES experiments are proportional to
the spectral weights of the low-energy QPs.\cite{ARPES} In
Ref.~\onlinecite{matsui03}, the spectral weights of a slightly
overdoped sample of Bi2223 were measured along the cut
$(\pi,0)\rightarrow (\pi/2,\pi/2)$; an almost constant total
spectral weight was reported in this experiment. It can be seen from
Fig.~\ref{fig:nFS} that the total spectral weight is approximately
constant along this cut, so the experimental result agrees with this
property of projected wave functions.

\hspace{.2cm}

An anisotropy of the ARPES intensity along the experimental FS (the
so-called nodal-anti\-nodal dichotomy) has been reported in a series
of experiments.\cite{zhou04,shen03} Experimentally, the spectral
weight measured in the anti-nodal region is suppressed in underdoped
compounds, while it is large in the optimally doped and overdoped
region. Usually, this effect is associated with formation of some
charge or spin order, static or fluctuating one. From
Fig.~\ref{fig:nFS} we see that a similar (but much weaker) tendency
can be observed in the framework of Gutzwiller-projected wave
functions. The experimentally observed effect is much stronger and a
claim that the nodal-antinodal dichotomy can be explained within
this framework would be too hasty.

\section{Superconducting order parameter}
\label{sec:sc}

\begin{figure}[h]
\includegraphics[scale=0.48,angle=0]{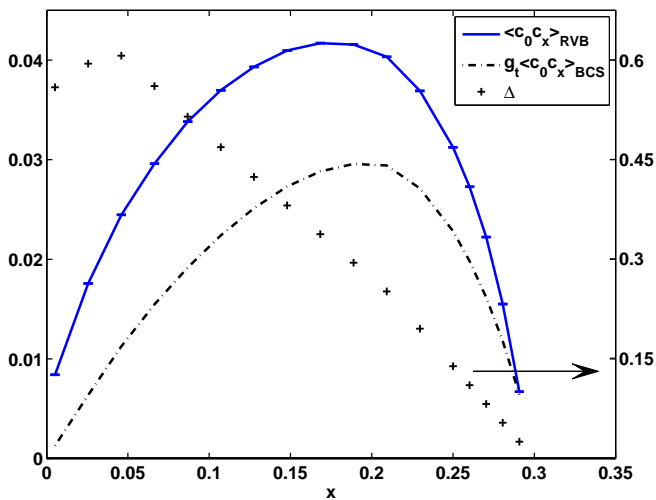}
\caption{(Color online) Doping dependency of the nearest-neighbor
superconducting order parameter $\Phi_{ij}$ (calculated in the
14$\times14$ system). The error bars are smaller than the symbol
size. The same quantity calculated in the Gutzwiller approximation
is also shown for comparison. The variational parameter $\Delta$ is
shown with the scale on the right.\label{fig:cc_x}}
\end{figure}

In Fig.~\ref{fig:cc_x}, we plot the nearest-neighbor superconducting
correlation $\Phi_{ij}$ (the Fourier transform of $\Phi_{\bm k}$
defined in (\ref{eq:cc})) as a function of doping. This curve shows
close quantitative agreement with the result of
Ref.~\onlinecite{paramekanti0103}, where the authors extracted the
superconducting order parameter from the long range asymptotics of
the nearest neighbor pairing correlator,
$\lim_{r\rightarrow\infty}\langle c_0 c_{\delta} c^{\dagger}_r
c_{r+\delta}^{\dagger}\rangle$. With the method employed here, we
find the same qualitative and quantitative conclusions of previous
authors:\cite{paramekanti0103,gros88} vanishing of superconductivity
at half filling $x\to 0$ and at the superconducting transition on
the overdoped side $x_c\simeq0.3$.\cite{note_cc}
 The optimal doping is near $x_{opt}\simeq0.18$. In the same plot we also show the commonly
used Gutzwiller approximation where the BCS order parameter is
renormalized by the factor $g_t = \frac{2 x}{1+x}$.\cite{rmft} The
Gutzwiller approximation underestimates the exact value by
approximately $25\%$.

\begin{figure}[!h]
\includegraphics[scale=0.65,angle=0]{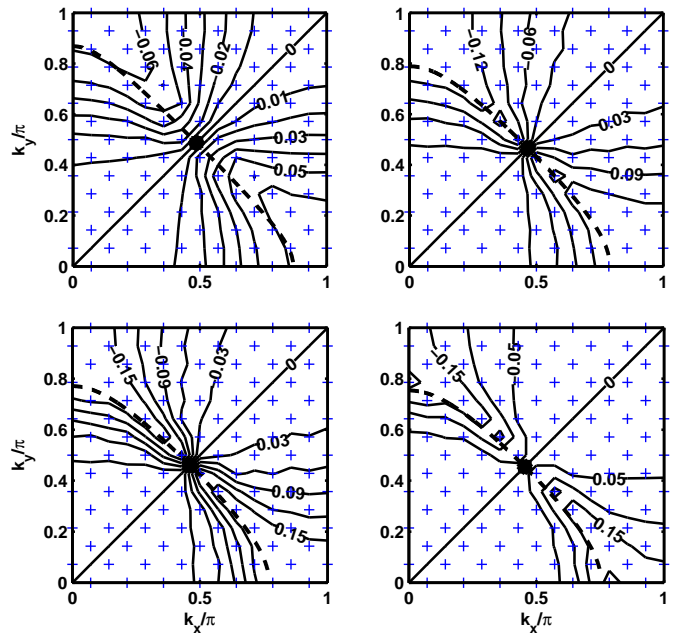}
\caption{(Color online) Pairing correlation $\Phi_{\bm k}$ in the
Brillouin Zone at differen doping levels: $x\simeq 3\%$ (upper
left), $11\%$ (upper right), $17\%$ (lower left) and $23\%$ (lower
right) in a $14\times14$ system with periodic-antiperiodic boundary
conditions. The dashed line is the unprojected FS.\label{fig:cc}}
\end{figure}

In Fig.~\ref{fig:cc}, we show contour plots of the superconducting
order parameter $\Phi_{\bm k}$ for four values of doping. It
resembles qualitatively the unprojected d-wave pairing amplitude,
but is somewhat distorted due to the particle-hole asymmetry (see
discussion in the previous and the following sections).

\section{Fermi surface}
\label{sec:FS}

In strongly interacting Fermi systems, the notion of a Fermi surface
(FS) is not at all clear. There are however several experimental
definitions of the FS. Most commonly, ${\bm k}_F$ is determined in
ARPES experiments as the maximum of $|{\bm \nabla}_{\bm k}n_{\bm
k}|$ or the locus of minimal gap along some cut in the $\bm
k$-plane.\cite{ARPES} The theoretically better defined locus of
$n_{\bm k} = 1/2$ is also sometimes used. The various definitions of
the FS usually agree within the experimental uncertainties.
Recently, the different definitions of the FS were theoretically
analyzed in Refs.~\onlinecite{gros06}~and~\onlinecite{sensarma06}.

In our present work, we propose an alternative definition of the
Fermi surface based on the ground state equal-time Green's
functions. In the unprojected BCS state, the underlying FS is
determined by the condition $|u_{\bm k}|^2 = |v_{\bm k}|^2$. We will
refer to this as the \textit{unprojected FS}. Since $|u_{\bm k}|^2$
and $|v_{\bm k}|^2$ are the residues of the QP poles in the BCS
theory, it is natural to replace them in the interacting case by
$Z^{+}_{\bm k}$ and $Z^{-}_{\bm k}$, respectively. We will therefore
define the \textit{effective FS} as the locus $Z^{+}_{\bm k} =
Z^{-}_{\bm k}$.

\begin{figure}[!h]
\includegraphics[scale=0.66,angle=0]{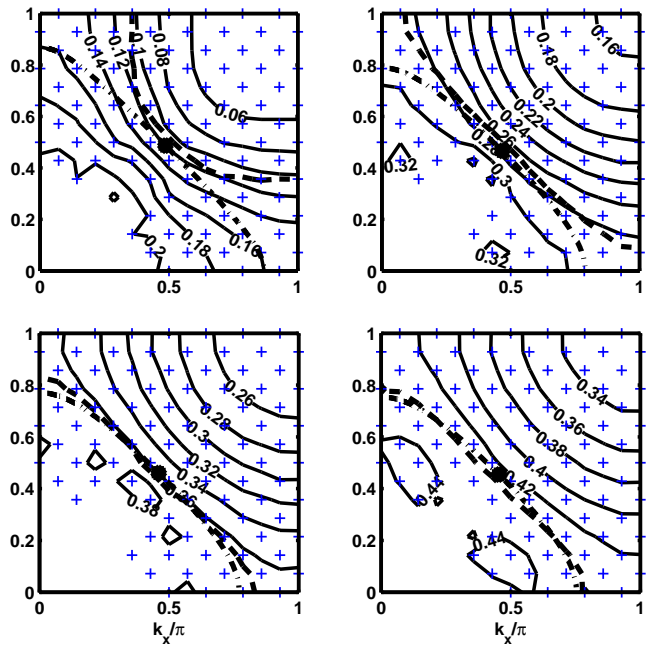}
\caption{(Color online) Contour plots of the total QP spectral
weight $Z_{\bm k}^{tot}$. The effective FS (dash line) and
unprojected FS (dash-dot line) are also shown. The doping levels are
$x\simeq 3\%$ (top left), $11\%$ (top right), $17\%$ (bottom left)
and $23\%$ (bottom right). The $+$ signs indicate points where the
values are known within small error bars.\label{fig:nFS}}
\end{figure}

In Fig.~\ref{fig:nFS} we plot the unprojected and the effective FS
which we obtained from VMC calculations. The contour plot of the
total QP weight is also shown. It is interesting to note the
following points:
\begin{itemize}
\item In the underdoped region, the effective FS is open and bent
outwards (hole-like FS). In the overdoped region, the effective FS
closes and embraces more and more the unprojected one with
increasing doping (electron-like FS).
\item Luttinger's rule\cite{luttinger61} is clearly violated in the
underdoped region, i.e.\ the area enclosed by the effective FS is
not conserved by the interaction; it is larger than that of the
unprojected one.
\item  In the optimally doped and overdoped region, the total spectral weight
is approximately constant along the effective FS within error bars.
In the highly underdoped region, we observe a small concentration of
the spectral weight around the nodal point ($\simeq 20\%$).
\end{itemize}

A large ``hole-like'' FS in underdoped cuprates has also been
reported in ARPES experiments by several
groups.\cite{ino04,yoshida03,shen03}

It should be noted that a negative next-nearest hopping $t'$ would
lead to a similar FS curvature as we find in the underdoped region.
We would like to emphasize that our original $t$--$J$ Hamiltonian as
well as the variational states do not contain any $t'$. Our results
show that the outward curvature of the FS is due to strong Coulomb
repulsion, without need of $t'$. The next-nearest hopping terms in
the microscopic description of the cuprates may not be necessary to
explain the FS topology found in ARPES experiments. Remarkably, if
the next-nearest hopping $t'$ is included in the variational ansatz
(and not in the original $t$--$J$ Hamiltonian), a finite and
negative $t'$ is generated, as it was shown in
Ref.~\onlinecite{himeda00}. Apparently, in this case the unprojected
FS has the tendency to adjust to the effective FS. A similar bending
of the FS was also reported in the recent analysis of the current
carried by Gutzwiller-projected QPs.\cite{nave06} A high-temperature
expansion of the momentum distribution function $n_{\bm k}$ of the
$t$-$J$ model was done in Ref.~\onlinecite{putikka98} where the
authors find a violation of Luttinger's rule and a negative
curvature of the FS. Our findings provide further evidence in this
direction.

A natural question is the role of superconductivity in the
unconventional bending of the FS. In the limit $\Delta\to 0$, the
variational states are Gutzwiller-projected excitations of the Fermi
sea and the spectral weights are step-functions at the (unprojected)
FS. In a recent paper\cite{yang06} it was shown that $\lim_{{\bm
k}\rightarrow {\bm k}^{+}_F} Z^{+}_{\bm k} = \lim_{{\bm
k}\rightarrow {\bm k}^{-}_F} Z^{-}_{\bm k}$ for the projected
Fermi-sea, which means that the unprojected and the effective FS
coincide in that case. This suggests that the ``hole-like'' FS
results from a non-trivial interplay between strong correlation and
superconductivity. At the moment, we lack a qualitative explanation
of this effect, however it may be a consequence of the proximity of
the system to the non-superconducting ``staggered-flux''
state\cite{lee00,ivanov03} or to
anti-ferromagnetism\cite{paramekanti0103,ivanov06} near
half-filling.



\hspace{.5cm}

\begin{acknowledgments}
We would like to thank George Jackeli for valuable input and
continuous support. We also thank Claudius Gros, Patrick Lee and
Seiji Yunoki for interesting discussions. During the final stage of
this work, we learned about a similar work by S.\ Yunoki which
complements and partly overlaps with the results presented
here.\cite{yunoki06} After the completion of this paper, we have
learned about a recent work by Chou et al.\ \cite{chou06} who also
reported relation~\eqref{eq:zpm}. This work was supported by the
Swiss National Science Foundation.
\end{acknowledgments}


\begin{thebibliography}{99}
\bibitem{note_norm}
As an alternative to the micro-canonical formulation in
Eqs.~\eqref{eq:wf} and \eqref{eq:excit}, one can work in the
grand-canonical ensemble, without the particle-number projector, but
with an additional fugacity factor.\cite{edegger05} In this paper,
the micro-canonical scheme is chosen for numerical convenience, as
it is commonly done in most VMC studies.
\bibitem{note_comp}
Our VMC results show good agreement with the hole spectral weight
reported in Ref.~\onlinecite{edegger06} where the authors used an
extended Gutzwiller approximation to calculate the same quantity in
the large-U Hubbard model. It should be noted, however, that the
asymmetry found in the present paper cannot be explained within any
known Gutzwiller approximation scheme. Our results are consistent
with other recent VMC calculations\cite{nave06, yang06, yunoki06}
and earlier calculations of $Z_{\bm k}^+$ in
Ref.~\onlinecite{paramekanti0103}.
\end{thebibliography}
\end{document}